\documentclass{article}
\usepackage{spconf,amsmath,graphicx}
\usepackage{multirow}
\usepackage{cite}
\usepackage{amssymb}

\title{Replay and Synthetic Speech Detection with Res2net Architecture}

\name{Xu Li$^{1}$\thanks{This work was done when Xu Li was an intern at Tencent AI Lab.}, Na Li$^2$, Chao Weng$^2$, Xunying Liu$^1$, Dan Su$^2$, Dong Yu$^2$, Helen Meng$^1$}
\address{
  $^1$The Chinese University of Hong Kong\\
  $^2$Tencent AI Lab, Tencent\\
  {\small \tt \{xuli, xyliu, hmmeng\}@se.cuhk.edu.hk, lina011779@126.com, \{cweng, dansu, dyu\}@tencent.com}}
  
\begin{document}
\ninept
\maketitle
\begin{abstract}
Existing approaches for replay and synthetic speech detection still lack generalizability to unseen spoofing attacks. This work proposes to leverage a novel model structure, so-called Res2Net, to improve the anti-spoofing countermeasure's generalizability. Res2Net mainly modifies the ResNet block to enable multiple feature scales. Specifically, it splits the feature maps within one block into multiple channel groups and designs a residual-like connection across different channel groups. Such connection increases the possible receptive fields, resulting in multiple feature scales. This multiple scaling mechanism significantly improves the countermeasure's generalizability to unseen spoofing attacks. It also decreases the model size compared to ResNet-based models. Experimental results show that the Res2Net model consistently outperforms ResNet34 and ResNet50 by a large margin in both physical access (PA) and logical access (LA) of the ASVspoof 2019 corpus. Moreover, integration with the squeeze-and-excitation (SE) block can further enhance performance. For feature engineering, we investigate the generalizability of Res2Net combined with different acoustic features, and observe that the constant-Q transform (CQT) achieves the most promising performance in both PA and LA scenarios. Our best single system outperforms other state-of-the-art single systems in both PA and LA of the ASVspoof 2019 corpus.

% is still limited when generalizing to unseen spoofing attacks. This work proposes to leverage Res2Net model architecture to improve the model's generalizability. Our experimental results illustrate a significant improvement of Res2Net models over ResNet34 and ResNet50. Moreover, we evaluate the generalizability of Res2Net combined with different acoustic features. Our best model is among the state-of-the-art single systems in both physical and logical access of ASVspoof 2019 corpus.

\end{abstract}

\section{INTRODUCTION}
\label{sec:intro}

Automatic speaker verification (ASV) systems aim at confirming a claimed speaker identity against a spoken utterance, which has been widely applied into commercial devices and authorization tools. However, it is also broadly noticed that malicious attacks can easily degrade a well-developed ASV system, and such attacks may be classified into impersonation \cite{wu2015spoofing}, replay \cite{wu2015spoofing}, voice conversion (VC) \cite{kinnunen2012vulnerability}, text-to-speech \cite{shchemelinin2013examining} synthesis (TTS) and the recently emerged adversarial attacks \cite{das2020attacker,li2020adversarial}.

Over the past decade, the speaker verification community has held several ASVspoof challenges \cite{wu2015asvspoof,kinnunen2017asvspoof,todisco2019asvspoof} to develop countermeasures mainly against replay, speech synthesis and voice conversion attacks. ASVspoof 2019 \cite{todisco2019asvspoof} is the latest one that includes all previous attacks within two sub-challenges: physical access (PA) and logical access (LA). PA considers spoofing attacks from replay while LA refers to attacks generated with TTS and VC techniques.

% The latest edition ASVspoof2019 \cite{todisco2019asvspoof} aims to address all previous attacks and further extend previous editions in three aspects: 1) update attacks with TTS and VC with state-of-the-art technologies; 2) create a more controlled setup for replay attacks, covering acoustic and microphone conditions and predefined replay device qualities; 3) adopt an evaluation metric to assess impacts of standalone anti-spoofing systems to a fixed ASV system. ASVspoof2019 consists of two sub-challenges: physical access (PA) and logical access (LA). PA considers spoofing attacks from replay while LA refers to attacks generated with TTS and VC.

A challenging issue for constructing reliable countermeasures is the defense against unseen attacks. This is fully considered in ASVspoof 2019 \cite{todisco2019asvspoof}, where the evaluation partition consists mostly of unseen spoofing attacks, generated with replay configurations and spoofing algorithms which differ from those in the training and development partitions. Countermeasures that perform well in the development set are still very likely to fail in the evaluation set, due to the lack of generalizability to unseen spoofing attacks \cite{todisco2019asvspoof}. This issue challenges the generalizability of developed countermeasures in face of unseen attack \cite{das2020assessing,li2020investigating}.
% This challenging issue seriously questions the generalization ability of developed countermeasures to deal with unseen attacks \cite{das2020assessing,li2020investigating}. 
To construct generalized countermeasures, existing efforts may be divided into three categories: feature engineering \cite{cheng2019replay,wu2012detecting}, system modeling \cite{alzantot2019deep,cai2019dku,lai2019assert} and loss criteria \cite{gomez2020kernel}. Among them, system modeling is most essential for data-driven countermeasures development with little human effort, and is the focus of this paper.

Many existing neural network architectures have been applied to designing powerful countermeasures against spoofing attacks, such as light convolutional neural network (LCNN) \cite{lavrentyeva2019stc,wu2020light}, residual neural network (ResNet) \cite{cai2019dku,lai2019assert} and their variations \cite{gomez2020kernel,cheng2019replay}. These models exhibit a strong ability in time and frequency domain modeling and achieve promising performance to capture the spoofing cues. However, when applied to unseen spoofing attacks, the performance of state-of-the-art (SOTA) systems is still limited \cite{todisco2019asvspoof}. ResNet and LCNN based systems mainly have two dimensions to control the model's capacity, i.e. width and depth. But simply increasing the width and depth is not efficient for improving the model's capacity. Specifically, since anti-spoofing countermeasures development requires high generalizability to unseen spoofing attacks, only increasing the width and depth can easily lead to over-fitting, due to the huge amount of parameters.
% Especially, for anti-spoofing countermeasures development which requires high generalizability to unseen spoofing attacks, simply increasing width and depth could easily fall into over-fitting due to huge amount of parameters.

In this work, we make the first attempt to leverage a novel Res2Net \cite{gao2019res2net} model architecture in anti-spoofing systems, motivated by their promising performance on vision tasks \cite{gao2019res2net,cao2019improved,zhou2020gfnet}. Res2Net focuses on the revision of the ResNet block to enable multiple feature scales. It splits the feature maps within one block into multiple channel groups and designs a residual-like connection across different channel groups. Such residual-like connection increases the possible receptive fields, resulting in multiple feature scales. This additional multi-scale reception improves the system's capacity and helps the system perform better when generalized to unseen spoofing attacks. Meanwhile, it also decreases the model size compared to traditional ResNet-based models. Our experiments verified a significant improvement of Res2Net50 over ResNet34 and ResNet50 in detecting both replay and synthetic speech audios. Moreover, integration with the squeeze-and-excitation (SE) block \cite{hu2018squeeze} can further enhance performance. For feature engineering, we investigate the generalizability of the Res2Net model that incorporates with different acoustic features, and observe that the constant-Q transform (CQT) achieves the most promising results in both PA and LA subsets of ASVspoof 2019. Our best single model outperforms the other SOTA single systems in both physical and logical scenarios.

\begin{table*}[ht]
\caption{The overall model architectures of ResNet34, ResNet50, Res2Net50 and SE-Res2Net50. The type of a residual block and the number of channels are specified inside the brackets, while the repeat times of each block on one stage are specified outside the brackets. ``2-d fc'' denotes a fully connected layer with 2 output units.}
\label{tab:overall-model-architecture}
\centering
\begin{tabular}{c|c|c|c|c}
         \hline
         \hline
         Stage & ResNet34 & ResNet50 & Res2Net50 & SE-Res2Net50 \\
         \hline
         \multirow{2}{*}{Conv1} & \multicolumn{2}{c|}{conv2d, $7 \times 7$, 16, $stride=2$} &  \multicolumn{2}{c}{\multirow{2}{*}{[conv2d, $3 \times 3$, 16, $stride=1$] $\times 3$}}\\
                                & \multicolumn{2}{c|}{max pool, $3 \times 3$, $stride=2$}  \\
         \hline
          Conv2 & [Basic BLK, $16$] $\times 3$ & [Bottleneck BLK, $16$] $\times 3$ & [Res2Net BLK, $16$] $\times 3$ & [SE-Res2Net BLK, $16$] $\times 3$ \\
          \hline
          Conv3 & [Basic BLK, $32$] $\times 4$ & [Bottleneck BLK, $32$] $\times 4$ & [Res2Net BLK, $32$] $\times 4$ & [SE-Res2Net BLK, $32$] $\times 4$ \\
          \hline
          Conv4 & [Basic BLK, $64$] $\times 6$ & [Bottleneck BLK, $64$] $\times 6$ & [Res2Net BLK, $64$] $\times 6$ & [SE-Res2Net BLK, $64$] $\times 6$ \\
          \hline
          Conv5 & [Basic BLK, $128$] $\times 3$ & [Bottleneck BLK, $128$] $\times 3$ & [Res2Net BLK, $128$] $\times 3$ & [SE-Res2Net BLK, $128$] $\times 3$ \\
          \hline
            \multicolumn{5}{c}{global average pool, 2-d fc, softmax} \\
         \hline
         \hline
\end{tabular}
\vspace{-0.3cm}
\end{table*}

The contributions of this work include: 1) Leveraging the Res2Net model architecture into anti-spoofing and verifying its significant improvements over ResNet34 and ResNet50 models; 2) Investigating the generalizability of the Res2Net model incorporated with different acoustic features in both PA and LA scenarios; 3) Developing a single model that outperforms other SOTA single systems for both PA and LA in ASVspoof 2019.
% The contribution of this work includes: 1) Incorporating Res2Net model architecture into anti-spoofing and evaluating its effectiveness over ResNet34 and ResNet50 models; 2) Evaluating the generalizability of Res2Net mdoel combined with different acoustic features; 3) Our best model is among the state-of-the-art single systems in both physical and logical scenarios.

The rest of this paper is organized as follows: Section~\ref{sec:approach} illustrates the Res2Net block and its integration with the SE block, followed by the overall model architectures. Experimental setup and results are discussed in Section~\ref{sec:expt-setup} and \ref{sec:expt-rst}, respectively. We conclude this work in Section~\ref{sec:conclusion}.

\section{Approach}
\label{sec:approach}

\subsection{Res2Net block}
\begin{figure}[ht]
    \centering
    \includegraphics[width=0.40\textwidth]{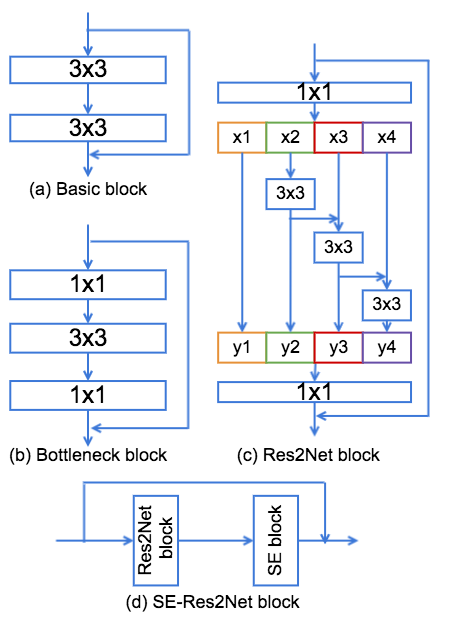}
    \caption{The illustration of the basic block, bottleneck block, Res2Net block (scale dimension $s=4$, the box in each color represents the feature maps within a channel group), and SE-Res2Net block.}
    \label{fig:basic_bottleneck_res2net_seres2net_block}
    % \vspace{-0.3cm}
\end{figure}

The Res2Net \cite{gao2019res2net} architecture aims at improving multi-scale representation by increasing the number of available receptive fields. This is achieved by connecting smaller filter groups within one block in a hierarchical residual-like style. The comparison among the basic block \cite{he2016deep}, bottleneck block \cite{he2016deep} and Res2Net block is illustrated in Fig.~\ref{fig:basic_bottleneck_res2net_seres2net_block}. The Res2Net block is modified from the bottleneck block. After a $1 \times 1$ convolution, it evenly splits the input feature maps by the channel dimension into $s$ subsets, denoted by $x_i$, where $i \in \{1,2,...,s\}$. Except for $x_1$, each $x_i$ is processed by a $3 \times 3$ convolutional filter $K_i()$. Starting from $i=3$, $x_i$ is added with the output of $K_{i-1}$ before being fed into $K_i()$. This process can be formulated as Eq.~\ref{eq:res2net-block}:
\begin{align}
\label{eq:res2net-block}
    y_i = \begin{cases}
     x_i, & i=1 \\
     K_{i}(x_i), & i=2 \\
     K_{i}(x_i+y_{i-1}), & 2 < i \leq s
    \end{cases}
\end{align}
where $s$ is defined as the scale dimension \cite{gao2019res2net}, indicating the number of partitions applied to split feature maps. This hierarchical residual-like connection enables multiple-size of receptive fields within one block, resulting in multiple feature scales. Finally, it concatenates all splits and passes them through a $1 \times 1$ convolution filter to maintain the channel size of this residual block.

Note that the hierarchical residual-like connection between filter groups also decreases the amount of model parameters. We assume that the feature maps after the first $1\times1$ convolution is $X \in \mathbb{R}^{I \times H_{in} \times W_{in}}$, and the one before the last $1\times1$ convolution is $Y \in \mathbb{R}^{O \times H_{out} \times W_{out}}$, where $H_{in}$, $W_{in}$, $H_{out}$ and $W_{out}$ are feature dimensions, and $I$ and $O$ are the number of channels for $X$ and $Y$, respectively. In the bottleneck block, it transforms $X$ to $Y$ by using a filter $W \in \mathbb{R}^{I \times O \times 3 \times 3}$ with a parameter size of $9 \times I \times O$. While in the Res2Net block, it splits $X$ into $s$ partitions with each $x_i \in \mathbb{R}^{\frac{I}{s} \times H_{in} \times W_{in}}$, for $i \in \{1, 2, ..., s\}$. Except for $x_{1}$, it transforms each $x_{i}$ into $y_{i} \in \mathbb{R}^{\frac{O}{s} \times H_{out} \times W_{out}}$ by using a filter $K_i \in \mathbb{R}^{\frac{I}{s} \times \frac{O}{s} \times 3 \times 3}$, resulting in a total parameter size of $\frac{9 \times I \times O \times (s-1)}{s^2}$, which is smaller than the parameter size of $W$.

\begin{table*}[th]
\caption{Results on the ASVspoof 2019 physical and logical access in terms of EER (\%) and t-DCF of different network architectures. (The input features for PA and LA are Spec and LFCC, respectively.)}

\label{tab:architecture-rst}
\centering
\begin{tabular}{c|c|c|c|c|c|c|c|c|c}
         \hline
         \hline
         \multirow{3}{*}{System} & \multirow{3}{*}{\# params} & \multicolumn{4}{|c}{Physical Access} & \multicolumn{4}{|c}{Logical Access}  \\
         \cline{3-10}
          & & \multicolumn{2}{|c}{Dev Set} & \multicolumn{2}{|c}{Eval Set} & \multicolumn{2}{|c}{Dev Set} & \multicolumn{2}{|c}{Eval Set} \\
         \cline{3-10}
          & & EER (\%) & t-DCF & EER (\%) & t-DCF & EER (\%) & t-DCF & EER (\%) & t-DCF \\
          \hline
          ResNet34 & 1.33M & 0.83 & 0.022 & 1.46 & 0.041 & 0.39 & 0.013 & 5.75 & 0.131 \\
          SE-ResNet34 & 1.34M & 0.57 & 0.015 & 1.32 & 0.037 & 0.35 & 0.011 & 4.69 & 0.103 \\
          \hline
          ResNet50 & 1.05M & 0.91 & 0.024 & 1.59 & 0.043 & 0.94 & 0.028 & 6.44 & 0.146 \\
          SE-ResNet50 & 1.09M & 0.70 & 0.020 & 1.37 & 0.038 & 0.47 & 0.008 & 5.06 & 0.109 \\
          \hline
             Res2Net50 & 0.88M & 0.45 & 0.012 & 0.91 & 0.026 & 0.36 & 0.011 & 4.55 & 0.099 \\
          SE-Res2Net50 & 0.92M & 0.52 & 0.012 & \textbf{0.74} & \textbf{0.021} & 0.23 & 0.005 & 2.87 & 0.079 \\
     Stat-SE-Res2Net50 & 0.96M & \textbf{0.35} & \textbf{0.001} & 1.00 & 0.027 & \textbf{0.20} & \textbf{0.004} & \textbf{2.86} & \textbf{0.068} \\
         \hline
         \hline
\end{tabular}
\vspace{-0.3cm}
\end{table*}

\subsection{Integration with the squeeze-and-excitation block}
The squeeze-and-excitation (SE) block \cite{hu2018squeeze} adaptively re-calibrates channel-wise feature responses by explicitly modeling the inter-dependencies between channels. This inter-dependencies modeling assigns different impact weights to channels, which improves the model's capacity to focus on channel information that is most related with spoofing cues. Motivated by this, we stack the Res2Net and SE blocks together to form the SE-Res2Net block, as shown in Fig.~\ref{fig:basic_bottleneck_res2net_seres2net_block} (d). Our experimental results demonstrated that this integration can bring further improvement for both PA and LA scenarios.

% \begin{table*}[ht]
% \caption{The configurations for backbone model architecture.}
% \label{tab:model-config}
% \centering
% \begin{tabular}{c|c|c|c|c}
%          \hline
%          \hline
%          Output size & ResNet34 & ResNet50 & Res2Net50 & SE-Res2Net50 \\
%          \hline
%          $129 \times 200$ & \multicolumn{4}{c}{conv, $7 \times 7$, 16, $stride=2$} \\
%          \hline
%          \multirow{2}{*}{$65 \times 100$} & \multicolumn{4}{c}{max pool, $3 \times 3$, $stride=2$} \\
%          \cline{2-5}
%           & [Basic BLK@$16$] $\times 3$ & [Bottleneck BLK@$16$] $\times 3$ & [Res2Net BLK@$16$] $\times 3$ & [SE-Res2Net BLK@$16$] $\times 3$ \\
%           \hline
%           $33 \times 50$ & [Basic BLK@$32$] $\times 4$ & [Bottleneck BLK@$32$] $\times 4$ & [Res2Net BLK@$32$] $\times 4$ & [SE-Res2Net BLK@$32$] $\times 4$ \\
%           \hline
%           $17 \times 25$ & [Basic BLK@$64$] $\times 6$ & [Bottleneck BLK@$64$] $\times 6$ & [Res2Net BLK@$64$] $\times 6$ & [SE-Res2Net BLK@$64$] $\times 6$ \\
%           \hline
%           $9 \times 13$ & [Basic BLK@$128$] $\times 3$ & [Bottleneck BLK@$128$] $\times 3$ & [Res2Net BLK@$128$] $\times 3$ & [SE-Res2Net BLK@$128$] $\times 3$ \\
%           \hline
%           $1 \times 1$ & \multicolumn{4}{c}{global average pool, 2-d fc, softmax} \\
%          \hline
%          \hline
% \end{tabular}
% \end{table*}

\subsection{Overall model architecture}
This work evaluates and compares the performances of ResNet34, ResNet50, Res2Net50 and SE-Res2Net50, and an overview of their architectures are shown in Table~\ref{tab:overall-model-architecture}. Notice that the Conv1 Stages of Res2Net50 and SE-Res2Net50 are slightly different from that of ResNet34 and ResNet50. The reason is that we adopt the updated ``v1b'' version of Res2Net provided by \cite{gao2019res2net}, which has been experimentally demonstrated to be more effective. The remaining parts for all models are identical, except for the block type. ResNet34, ResNet50, Res2Net50 and SE-Res2Net50 adopt the basic, bottleneck, Res2Net and SE-Res2Net blocks, respectively. Based on our experiments, we decrease the block expansion in ResNet50, Res2Net50 and SE-Res2Net50 from 4 to 2, to reduce the number of model parameters and prevent over-fitting issues. The scale dimension $s$ in Res2Net-based models is experimentally set as 4.

\section{EXPERIMENTAL SETUP}
\label{sec:expt-setup}

\textbf{Dataset:} All experiments are conducted under the ASVspoof 2019 challenge, which contains two subset evaluations: PA and LA. The detailed description of these two subsets are shown in Table~\ref{tab:asvspoof2019-corpus}. For each subset, our models are trained on the training partition, and the development partition is used for model selection. The evaluation partition consists mostly of unseen spoofing attacks, generated with either replay configurations or spoofing algorithms which differ from those in the training and development partitions. Systems are evaluated by the tandem detection cost function (t-DCF) \cite{todisco2019asvspoof} and equal error rate (EER) \cite{todisco2019asvspoof}. The log-probability of the bonafide class is adopted as the score for t-DCF and EER computation.

\textbf{Feature extraction:} Three acoustic features are evaluated in our experiments: log power magnitude spectrogram (Spec), linear frequency cepstral coefficients (LFCC) and constant-Q transform (CQT). The Spec is extracted with a ``Hanning'' window having size of 25ms and step-size of 10ms, and 512 FFT points are applied. The LFCC exactly follows the official baseline provided in the ASVspoof 2019 \cite{todisco2019asvspoof}, extracted with 20ms window length, 512 FFT points and 20 filters with their delta and double delta coefficients, making 60-dimensional feature vectors. The CQT is extracted with 16ms step size, Hanning window, 9 octaves with 48 bins per octave. All features are truncated along the time axis to reserve exactly 400 frames. The feature less than 400 frames would be extended by repeating their contents.
\begin{table}[ht]
\caption{Summary of the ASVspoof2019 corpus}
\label{tab:asvspoof2019-corpus}
\centering
\begin{tabular}{c|c|c|c|c}
         \hline
         \hline
         \multirow{2}{*}{Partition} & \multicolumn{2}{|c}{Physical Access} & \multicolumn{2}{|c}{Logical Access} \\
         \cline{2-5}
         & \#Bonafide & \#Spoofed & \#Bonafide & \#Spoofed \\
         \hline
        Train & 5,400 & 48,600 & 2,580 & 22,800 \\
         Dev. & 5,400 & 24,300 & 2,548 & 22,296 \\
        Eval. & 18,090 & 116,640 & 7,355 & 63,882 \\
         \hline
         \hline
\end{tabular}
\vspace{-0.3cm}
\end{table}

\textbf{Training strategy:} 
% In our experiments, we compare the performance of Res2Net50 to both ResNet34 and ResNet50 systems. All three models share the same configurations of the backbone model architecture, as shown in Table~\ref{tab:model-config}, except for the block type. ResNet34, ResNet50 and Res2Net50 adopts basic, bottleneck and Res2Net module, respectively. The block expansion of ResNet50 and Res2Net50 is decreased from 4 to 2, to reduce model parameters and prevent over-fitting issues. The scale dimension $s$ in Res2Net50 is set as 4. 
The binary cross entropy is used as the loss function to train all models. Adam \cite{kingma2014adam} is adopted as the optimizer with $\beta_1=0.9$, $\beta_2=0.98$ and weigth decay $10^{-9}$. The learning rate scheduler increases the learning rate linearly for the first 1000 warm-up steps and then decreases it proportionally to the inverse square root of the step number \cite{vaswani2017attention}. All models are trained with 20 epochs, and the model with lowest EER on development set is chosen to be evaluated. Our codes have been made open-source\footnote{https://github.com/lixucuhk/ASV-anti-spoofing-with-Res2Net}.

% \begin{table}[ht]
% \caption{The configurations for backbone model architecture.}
% \label{tab:model-config}
% \centering
% \begin{tabular}{c|c|c|c|c}
%          \hline
%          \hline
%          Config. & Block1 & Block2 & Block3 & Block4 \\
%          \hline
%          \#blocks & 3 & 4 & 6 & 3 \\
%          \#channels & 16 & 32 & 64 & 128 \\
%          \hline
%          \hline
% \end{tabular}
% \end{table}

\section{Results}
\label{sec:expt-rst}

\subsection{Effectiveness of the Res2Net architecture}
This section evaluates the effectiveness of the proposed Res2Net-based architectures for detecting spoofing samples. We leverage the results of ResNet34-based and ResNet50-based models for comparison, as shown in Table~\ref{tab:architecture-rst}. The input features for PA and LA evaluation are Spec and LFCC, respectively. Upon comparing ResNet34 and ResNet50, we observe that ResNet34 outperforms ResNet50 in all conditions, which indicates that simply increasing the depth is not efficient for generalizability enhancement.
% The parameter size of ResNet50 is smaller than ResNet34, because we decreased the expansion in ResNet50 from 4 to 2. 
% We also conducted experiments on ResNet50 with the expansion being 4, and observed worse performance.
After involving Res2Net50 into comparison, we observe that it significantly outperforms both ResNet34 and ResNet50. Specifically, Res2Net50 respectively outperforms ResNet34 and ResNet50 by a relative EER reduction of 37.7\% and 42.8\% on the PA evaluation set, and 20.9\% and 29.3\% on the LA evaluation set. Similar gains are also observed under the t-DCF metric. We also observe that when compared to ResNet34 and ResNet50, the model size of Res2Net50 is relatively reduced by 33.8\% and 16.2\%, respectively, which verifies the efficiency of the Res2Net architecture in detecting spoofing attacks.
% We observe that when compared to ResNet34 and ResNet50, the model size of Res2Net50 is relatively reduced by 33.8\% and 16.2\%, respectively, while the performance still has a large gain. Specifically, Res2Net50 respectively outperforms ResNet34 and ResNet50 by a relative EER reduction of 37.7\% and 42.8\% on PA evaluation set, and 20.9\% and 29.3\% on LA evaluation set. Similar gains are also observed under the t-DCF metric. 

\begin{table}[ht]
\caption{Results on the ASVspoof 2019 physical access in terms of EER (\%) and t-DCF of SE-Res2Net50 with different input features.}
\label{tab:feature-rst-pa}
\centering
\begin{tabular}{c|c|c|c|c}
         \hline
         \hline
         \multirow{2}{*}{Features} & \multicolumn{2}{|c}{Dev Set} & \multicolumn{2}{|c}{Eval Set} \\
         \cline{2-5}
         & EER (\%) & t-DCF & EER (\%) & t-DCF \\
         \hline
         Spec & 0.519 & 0.0120 & 0.741 & 0.0207 \\
         LFCC & 0.833 & 0.0229 & 1.465 & 0.0434 \\
         CQT  & 0.329 & 0.0086 & 0.459 & 0.0116 \\
         \hline
         fusion & \textbf{0.096} & \textbf{0.0028} & \textbf{0.287} & \textbf{0.0075} \\
         \hline
         \hline
         \multicolumn{5}{c}{ASVspoof 2019 Baseline \cite{todisco2019asvspoof}} \\
         \hline
         CQCC-GMM & 9.87 & 0.1953 & 11.04 & 0.2454 \\
         LFCC-GMM & 11.96 & 0.2554 & 13.54 & 0.3017 \\
         \hline
         \hline
\end{tabular}
\vspace{-0.3cm}
\end{table}

\begin{table}[ht]
\caption{Results on the ASVspoof 2019 logical access in terms of EER (\%) and t-DCF of SE-Res2Net50 with different input features.}
\label{tab:feature-rst-la}
\centering
\begin{tabular}{c|c|c|c|c}
         \hline
         \hline
         \multirow{2}{*}{Features} & \multicolumn{2}{|c}{Dev Set} & \multicolumn{2}{|c}{Eval Set} \\
         \cline{2-5}
         & EER (\%) & t-DCF & EER (\%) & t-DCF \\
         \hline
         Spec & 0 & 0 & 8.783 & 0.2237 \\
         LFCC & 0.228 & 0.0051 & 2.869 & 0.0786 \\
         CQT & 0.432 & 0.0143 & 2.502 & 0.0743 \\
         \hline
         fusion & \textbf{0} & \textbf{0} & \textbf{1.892} & \textbf{0.0452} \\
         \hline
         \hline
         \multicolumn{5}{c}{ASVspoof 2019 Baseline \cite{todisco2019asvspoof}} \\
         \hline
         CQCC-GMM & 0.43 & 0.0123 & 9.57 & 0.2366 \\
         LFCC-GMM & 2.71 & 0.0663 & 8.09 & 0.2116 \\
         \hline
         \hline
\end{tabular}
\vspace{-0.3cm}
\end{table}

Furthermore, integration with the SE block further improves the performance for all model architectures. For Res2Net50, the SE block achieves a relative EER reduction of 18.7\% and 36.9\% for PA and LA evaluation set, respectively. For horizontal comparisons, SE-Res2Net50 outperforms SE-ResNet34 and SE-ResNet50 by a relative EER reduction of 43.9\% and 46.0\% respectively on the PA evaluation set, and 38.8\% and 43.3\% respectively on the LA evaluation set. We also observe similar gains under the t-DCF metric. 

A further attempt replaces the average pooling (AvgPool) layer in SE-Res2Net50 by the statistics pooling (StatPool) layer, the performance of which is shown at the bottom in Table~\ref{tab:architecture-rst}. We observe that the StatPool layer improve the performance on development set for both PA and LA scenarios, but less improvement is observed on the LA evaluation set and performance decline is observed on the PA evaluation set. This suggests that the StatPool layer can lead to over-fitting issues for ASV anti-spoofing tasks.
% but performance has less improvement on LA evaluation set and even gets worse on PA evaluation set. This suggests that the StatPool layer has no significant benefits for model's generalization to unseen attacks.

\subsection{Feature engineering}
This section evaluates different acoustic features incorporated with SE-Res2Net50. The results of PA and LA scenarios are shown in Table~\ref{tab:feature-rst-pa} and Table~\ref{tab:feature-rst-la}, respectively. We have different observations in PA and LA scenarios. For instance, the Spec performs relatively well in the PA scenario, while the performance is much worse compared to the other two features in the LA scenario. Actually, we also observe limited generalization in the LA scenario for the Spec feature integrated with SE-ResNet34 and SE-ResNet50. One possible reason is that LA attacks are synthetic speech where the phase information is fake and estimated during the synthesis process. Hence, most spoofing cues for LA attacks exist in the phase information of the audio. While the Spec features we adopted are the log magnitudes of the spectrum, which contain no phase information, resulting in poor generalization for LA attacks. The LFCC is more suitable in detecting LA attacks than PA attacks. Interestingly, the CQT performs the best in both scenarios. Moreover, we also conduct experiments on other features, such as filter-banks and constant-Q cepstral coefficients (CQCC). But we observe similar or worse performance than the three features above.

\begin{table}[th]\scriptsize
\caption{Performance comparison of our proposed system to some known state-of-the-art single systems on the ASVspoof 2019 PA and LA evaluation set. Some systems reported results on only one access, and we denote the absent results as ``--'' in the table.}
\label{tab:system-comparison}
\centering
\begin{tabular}{c|c|c|c|c}
         \hline
         \hline
         \multirow{2}{*}{System} & \multicolumn{2}{|c}{Physical Access} & \multicolumn{2}{|c}{Logical Access} \\
         \cline{2-5}
         & EER (\%) & t-DCF & EER (\%) & t-DCF \\
         \hline
         Spec+ResNet+CE\cite{alzantot2019deep} & 3.81 & 0.0994 & 9.68 & 0.2741 \\
         MFCC+ResNet+CE\cite{alzantot2019deep} &   -- &     -- & 9.33 & 0.2042 \\
         CQCC+ResNet+CE\cite{alzantot2019deep} & 4.43 & 0.1070 & 7.69 & 0.2166 \\
          Spec+ResNet+CE\cite{lai2019assert} & 1.29 & 0.036 & 11.75 & 0.216 \\
           Joint-gram+ResNet+CE\cite{cai2019dku} & 1.23 & 0.0305 & -- & -- \\
           GD-gram+ResNet+CE\cite{cai2019dku} & 1.08 & 0.0282 & -- & -- \\
          LFCC+LCNN+A-softmax\cite{lavrentyeva2019stc} & 4.60 & 0.1053 & 5.06 & 0.1000 \\
           FFT+LCNN+A-softmax\cite{lavrentyeva2019stc} &    -- &     -- & 4.53 & 0.1028 \\
           CQT+LCNN+A-softmax\cite{lavrentyeva2019stc} &  1.23 & 0.0295 &   -- & -- \\
        FG-CQT+LCNN+CE\cite{wu2020light} & -- & -- & 4.07 & 0.102 \\
        Spec+LCGRNN+GKDE-Softmax\cite{gomez2020kernel} & 1.06 & 0.0222 & 3.77 & 0.0842 \\
        % Spec+LC-GRNN+GKDE-Soft.\&Cont.\cite{gomez2020kernel} & 0.97 & 0.0210 & 3.39 & 0.0805 \\
        Spec+LCGRNN+GKDE-Triplet\cite{gomez2020kernel} & 0.92 & 0.0198 & 3.03 & 0.0776 \\
               MGD+ResNeWt+CE\cite{cheng2019replay} & 2.15 & 0.0465 & -- & -- \\
            CQTMGD+ResNeWt+CE\cite{cheng2019replay} & 0.94 & 0.0250 & -- & -- \\
    Fbanks\&CQT+ResNeWt+CE\cite{cheng2019replay} & 0.52 & 0.0134 & -- & -- \\
         \hline
         \textbf{Ours: CQT+SE-Res2Net50+CE} & \textbf{0.459} & \textbf{0.0116} & \textbf{2.502} & \textbf{0.0743} \\
         \hline
         \hline
\end{tabular}
\vspace{-0.3cm}
\end{table}

When compared to the baseline systems \cite{todisco2019asvspoof} provided by ASVspoof2019, the CQT-based system achieves a relative EER reduction over 95.8\% in the PA evaluation set, and a relative EER reduction over 69.1\% in the LA evaluation set. To take advantage of the complementarity across these three features, we designed a fusion system operated by simply averaging the output scores of three systems. It achieves an EER of 0.287\% and t-DCF of 0.0075 in the PA evaluation set, and an EER of 1.892\% and t-DCF of 0.0452 in the LA evaluation set.

\subsection{Comparison with the state-of-the-art single systems}
This section compares our best single system, i.e. SE-Res2Net50 incorporated with the CQT, with some of the reported SOTA single systems on the ASVspoof2019 PA and LA evaluation sets. 
Some of the top single systems are shown in Table~\ref{tab:system-comparison}, according to our best knowledge.
% We list some of the top single systems according to our best knowledge, as shown in Table~\ref{tab:system-comparison}. 
The systems are denoted by a name consisting of input features, system architecture and loss criteria. 

We observe that only a few systems achieve either an EER below 1.0\% on the PA evaluation set, or an EER below 4.0\% on the LA evaluation set, and rarely do systems have promising performance on both, which indicates the challenge in detecting unseen spoofing attacks. Most of the well-performing systems explore powerful model architectures and loss criteria. For PA attacks, the system in \cite{cheng2019replay} has very promising performance and achieved the top single system in the PA scenario of the ASVspoof2019 competition. This is outperformed by our system at a relative EER reduction of 11.5\%, depicting the effectiveness of our proposed system. For LA attacks, our system also outperforms other SOTA systems by a large margin.
% In the PA scenario, we observe that only a few systems achieve an EER below 1.0\% on the evaluation set, which indicates a challenging issue of defense against unseen spoofing attacks. Most of the well-performing systems dedicate efforts in developing powerful model architectures and loss criterion. Among them, the system \cite{cheng2019replay} has promising performance and achieved the top1 single system in the PA scenario of the ASVspoof2019 competition. Our system still outperforms it by a relative EER reduction of 11.5\%, depicting its effectiveness of detecting spoofing attacks. In the LA scenario, our system also outperforms the other state-of-the-art systems by a large performance gain.

\section{CONCLUSION}
\label{sec:conclusion}
This work proposes to leverage the Res2Net architecture into ASV anti-spoofing systems. Res2Net revises the ResNet block to enable multiple feature scales, which effectively improves the countermeasure's generalizability to unseen spoofing attaks. Experimental results show significant improvement of Res2Net50 over the ResNet34 and ResNet50 systems in both PA and LA scenarios. Moreover, we incorporated different acoustic features into Res2Net50, and found that the CQT achieves the most promising results for both PA and LA attacks. Our best single model outperforms other SOTA single systems in both PA and LA scenarios. Future work will explore loss criteria with better generalization ability to deal with ASV anti-spoofing tasks.

\section{ACKNOWLEDGEMENT}
This work is partially supported by HKSAR Government’s Research Grants Council General Research Fund (Project No. 14208718).

% \section{REFERENCES}
% \label{sec:refs}

\bibliographystyle{IEEEbib}
\bibliography{strings,refs}

\end{document}